\documentclass[aps,prl,reprint,preprintnumbers,showpacs,groupedaddress,amsmath,amssymb,floatfix]{revtex4-1}

\usepackage{graphicx,epsfig,dcolumn,multirow,subcaption}
\usepackage{natbib}
\usepackage[normalem]{ulem}
\newcolumntype{d}[1]{D{.}{.}{#1}}

\RequirePackage{xspace}

\newcommand{\simgt}{\,\hbox{\lower0.6ex\hbox{$\sim$}\llap{\raise0.6ex\hbox{$>$}}}\,}
\newcommand{\simlt}{\,\hbox{\lower0.6ex\hbox{$\sim$}\llap{\raise0.6ex\hbox{$<$}}}\,}

\newcommand{\ket}[1]{\ensuremath{| #1 \rangle }}
\newcommand{\CP}{\ensuremath{C\!P}\ }
\newcommand{\lt}{\left}
\newcommand{\rt}{\right}

\newcommand{\real}{\mathrm{Re}\,}
\newcommand{\dm}{\ensuremath{\Delta M}}
\newcommand{\dg}{\ensuremath{\Delta \Gamma}}
\newcommand{\ov}[1]{\overline{#1}}

\newcommand{\braOket}[3]{\langle#1|#2|#3\rangle}

\newcommand{\eq}[1]{Eq.~(\ref{#1})}
\newcommand{\eqsand}[2]{Eqs.~(\ref{#1}) and (\ref{#2})}
\newcommand{\fig}[1]{Fig.~\ref{#1}}
\newcommand{\bb}{\ensuremath{B\!-\!\Bbar{}\,}}

\newcommand{\bbmd}{\bbd\ mixing}

\newcommand{\bbm}{\bb\ mixing}

\newcommand{\bbd}{\ensuremath{B_d\!-\!\Bbar{}_d\,}}
\newcommand{\bbs}{\ensuremath{B_s\!-\!\Bbar{}_s\,}}
\newcommand{\bbq}{\ensuremath{B_q\!-\!\Bbar{}_q\,}}
\newcommand{\Bbar}{\,\overline{\!B}}

\newcommand{\gev}{\ensuremath{\, \mathrm{GeV}}}
\newcommand{\tev}{\ensuremath{\, \mathrm{TeV}}}

\newcommand{\lqcd}{\Lambda_{\rm QCD}}

\newcommand{\nn}{\nonumber\\}

\usepackage{pstricks}
\newcommand{\uli}[1]{{#1}}

\begin{document}

\preprint{TTP15-006}

\title{
\boldmath
Penguin contributions to \CP phases in $B_{d,s}$ decays to charmonium
\unboldmath}

\author{Philipp Frings$^{\,a}$,
  Ulrich Nierste$^{\,a}$, and Martin Wiebusch$^{\,c}$
\vspace{0.6cm}}

\affiliation{
\mbox{$^{a}$ Institut f\"ur Theoretische Teilchenphysik,
Karlsruhe Institute of Technology, D-76128 Karlsruhe, Germany,}
\mbox{email: philipp.frings@kit.edu, ulrich.nierste@kit.edu}\\
\mbox{$^{c}$  IPPP, Department of Physics, University of Durham, DH1
	      3LE, UK,	{e-mail: martin.wiebusch@durham.ac.uk}}
}

\begin{abstract}
  The precision of the \CP phases $2\beta$ and $2\beta_s$ determined
  from the mixing-induced \CP asymmetries in $B_d\to J/\psi K_S$ and
  $B_s\to J/\psi \phi$, respectively, is limited by the unknown
  long-distance contribution of a penguin diagram involving up
  quarks. {The penguin contribution is expected to be comparable
    in size to the precision of the LHCb and Belle II
    experiments and therefore limits the sensitivity of the measured
    quantities to new physics.}
  We analyze the infrared QCD structure of this contribution and
  find that all soft and collinear divergences either cancel between
  different diagrams or factorize into matrix elements of local
  four-quark operators {up to terms suppressed by $\lqcd/m_{\psi}$,
    where $m_{\psi}$ denotes the $J/\psi$ mass}. Our results,
  {which are based on an operator product expansion,} allow us to
  calculate the penguin-to-tree ratio $P/T$ in terms of the matrix
  elements of these operators and to constrain the penguin contribution
  to the phase $2\beta$ as $|\Delta \phi_d|\leq {0.68}^\circ$. The
  penguin contribution to $2\beta_s$ is bounded as $|\Delta
  \phi_s^{0}|\leq {0.97}^\circ$, $|\Delta \phi_s^{\parallel}|\leq
  {1.22}^\circ$, and $|\Delta \phi_s^{\perp}|\leq {0.99}^\circ$
  for the {case of} longitudinal, parallel, and perpendicular $\phi$
  and $J/\psi$ polarizations, respectively. {We further place bounds on
    $|\Delta \phi_d|$ for $B_d\to \psi(2S) K_S$ and the polarization
    amplitudes in $B_d\to J/\psi K^*$.}  In our approach it is further
  possible to constrain $P/T$ for decays in which $P/T$ is
  Cabibbo-unsuppressed and we {derive upper limits on} the 
  {penguin contribution to the} mixing-induced \CP asymmetries
  {in} $B_d\to J/\psi \pi^0$, $B_d\to J/\psi \rho^0$, $B_s\to J/\psi
  K_S$, and {$B_s\to J/\psi K^*$. For all studied decay modes we
    also constrain the sizes of the direct \CP asymmetries.}  
\end{abstract}


\maketitle

\section{Introduction}
The mixing-induced \CP asymmetry in the decay $B_d\to J/\psi K_S$ is the
key quantity to measure the \CP phase of the \bbmd\ amplitude.  Within
the Standard Model (SM) this \CP asymmetry {$A_{\rm CP}^{B_{d}\to
    J/\psi K_S }(t)$} determines the angle
$\beta=\arg[-V_{tb}V_{td}^*/(V_{cb}V_{cd}^*)]$ of the unitarity
triangle. {The $B$ factories BaBar and Belle had been designed to
  measure $A_{\rm CP}^{B_{d}\to J/\psi K_S }(t)$ to a high precision to
  probe the Kobayashi-Maskawa (KM) mechanism of CP violation.  Within
  the Standard Model, the KM phase is the only source of CP violation in
  weak transitions and therefore must correctly describe \emph{all}\ CP
  asymmetries measured in weak hadron decays.  The measurement of
  $\beta$ at the $B$ factories gave us sufficient confidence that the KM
  mechanism correctly describes CP violation in both $K$ and $B_d$
  decays and \uli{led} to the dedication of the 2008 Nobel Prize in Physics
  to Makoto Kobayashi and Toshihide Maskawa.  Today's focus of flavor
  physics is the search for physics beyond the Standard Model which
  reveals itself in small deviations from the KM picture. In generic
  models of new physics \bbm\ probes new physics associated with scales
  beyond 100\tev; reducing the uncertainties of Standard-Model
  predictions is therefore of utmost importance.}  $B_s\to J/\psi
\phi$ 
{is the analogous key mode in} the \bbs\ system. Since the unitarity
of the Cabibbo-Kobayashi-Maskawa (CKM) matrix essentially fixes
$\beta_s=\arg[-V_{tb}^*V_{ts}/(V_{cb}V_{cs}^*)]=1.0^\circ$ {to a
  very small value}, \CP studies of $B_s\to J/\psi \phi$ directly probe
physics beyond the SM. The decay amplitude $A_f$ for an $\ov b\to \ov c
c \ov s$ decay $B_q \to f$, where $f$ is a \CP eigenstate consisting of
a charmonium state and a light meson, can be written as
\begin{equation}
A_f = \lambda_c^s T_f + \lambda_u^s P_f \label{eq:atp}
\end{equation}
with $\lambda_{p}^s=V_{pb}^*V_{ps}$, {$p=u,c$,} and
\begin{eqnarray}
 T_f &=& \frac{G_F}{\sqrt{2}} \braOket{f}{
     C_1 Q_1^c +C_2 Q_2^c \uli{+} \sum_{j} C_j Q_j}{B_q},\\
 P_f &=& \frac{G_F}{\sqrt{2}} \braOket{f}{
     C_1 Q_1^u +C_2 Q_2^u \uli{+} \sum_{j} C_j Q_j}{B_q}.
\label{eq:tp}
\end{eqnarray}
Here,  $Q_1=\ov s{}^\alpha \gamma_\mu (1-\gamma_5) q^\beta
\ov q{}^\beta \gamma^\mu (1-\gamma_5)b^\alpha$ and $Q_2=\ov s{}^\alpha
\gamma_\mu (1-\gamma_5) q^\alpha
\ov q{}^\beta \gamma^\mu (1-\gamma_5)b^\beta$ are the current-current
operators. The index $j$ labels the penguin operators $Q_j$ which involve the 
CKM elements $\lambda^s_t =-\lambda^s_c-\lambda^s_u$. While the QCD penguin 
operators 
$Q_{3-6}$ and $Q_{8G}$ are important for this paper (see
Ref.~\cite{bbl} for their definition), electroweak penguin operators
have negligible effects.  The time-dependent \CP asymmetry
$A_{\rm CP}^{B_q\to f}(t)\equiv
[\Gamma(\Bbar_q(t)\to f)-\Gamma(B_q(t)\to f)]/[\Gamma(\Bbar_q(t)\to f)+
\Gamma(B_q(t)\to f)]$ reads
\begin{equation}
A_{\rm CP}^{B_q\to f}(t) =
\frac{S_f \sin(\dm_q t)- C_f \cos(\dm_q t)}{\cosh(\dg_q
  t/2)+A_{\dg_q}\sinh(\dg_q t/2)}.
\end{equation}
Here $\dm_q$ and $\dg_q$ are the mass and width difference,
respectively, between the mass eigenstates of the \bbq\ system.  We
write $S_f{\approx -\eta_f}\sin(\phi_q+ \Delta \phi_q)$, where 
{$\CP\!\ket f=\eta_f\ket f$ and } $\phi_q$
is the \CP phase in the limit $P_f=0$. The SM predictions are
$\phi_d=2\beta$ and $\phi_s=-2\beta_s$. To first order in
$\epsilon=|V_{us}V_{ub}/(V_{cs}V_{cb})|\approx 0.02$ one has
\begin{equation}
    \tan(\Delta \phi)\simeq 2
    \epsilon\sin \gamma \, \real \frac{P_f}{T_f} .
	\label{eq:Dphi}
\end{equation}
Comparing \eq{eq:Dphi} (with $\gamma = (69.7\pm2.8)^\circ $)
with the present experimental world average $\sin \phi_d=
  0.679 \pm 0.020$ \cite{hfag} (meaning an error of $1.6^\circ$ for
  $\phi_d$) shows that the penguin contribution already matters now and
  will certainly do so for future measurements at LHCb and Belle II.
$T_f$ and $P_f$ are non-perturbative multi-scale matrix elements, which
defy calculations from first principles of QCD.

For the prediction of the branching ratio $B(B_d\to J/\psi K_S)$ one
only needs $T_f$, which was addressed with the method of QCD
factorization \cite{qcdf} in Ref.~\cite{Chay:2000xn}: in the limit of
infinite charm and bottom masses $T_f$ can be expressed in terms of the
$J/\psi$ decay constant and the $B_d\to K_S$ form factor. The result of
Ref.~\cite{Chay:2000xn} underestimates $B(B_d\to J/\psi K_S)$ by a
factor of 8. This failure, however, is not surprising, because the
corrections to the infinite-mass limit are of order
$\lqcd/(m_c\alpha_s)$ and therefore numerically unsuppressed for the
actual value of the charm mass \cite{qcdf,Beneke:2002bs}.  The standard
approach to quantify $P_f/T_f$ in $B_d\to J/\psi K_S$ uses the
approximate SU(3)$_{\rm F}$ symmetry of QCD (or its U-spin subgroup)
which relates the decay of interest to $b\to c \ov c d$ modes like
$B_s\to J/\psi K_S$ and $B_d \to J/\psi \pi^0$
{\cite{Fleischer:1999nz,su3}}.  A drawback of this method is our poor
knowledge of the quality of the SU(3)$_{\rm F}$ symmetry in $B_{d,s}\to
J/\psi X$ (with $X=K_S,\pi^0,\ldots$) decays. (Comparisons of branching
ratios essentially test SU(3)$_{\rm F}$ in $T_f$ only, with little
sensitivity to $P_f$.) Furthermore, the $b\to c\ov c d$ control channels
have 20 times smaller statistics than their $b\to c\ov c s$
counterparts.  SU(3)$_{\rm F}$ seemingly fails in $B_s\to J/\psi \phi$,
because the $\phi$ meson cannot be closely approximated by an
SU(3)$_{\rm F}$ eigenstate, but is an equal mixture of octet and
singlet.

In this paper, we present a dynamical calculation of $P_f/T_f$ which does
not assume an approximate SU(3)$_{\rm F}$ symmetry. Our results permit,
for the first time, the prediction of $S_f$ and $C_f$ also for $b\to
c\ov c d$ decays.

\section{Operator product expansion}
For definiteness we first specify the discussion to $B_d\to J/\psi K_S$
and return to $B_s\to J/\psi \phi$ and other modes in the phenomenology
section.  {For} $B(B_d\to J/\psi K_S)$ we only need $T_f$ and can
  neglect the penguin coefficients.  It is useful to express $T_f$ in
terms of the matrix elements of
\begin{eqnarray}
Q_{0V} & \equiv & \ov s{} \gamma_\mu (1-\gamma_5) b \,
		  \ov c{}  \gamma^\mu c ,
		\nn
Q_{0A} & \equiv & \ov s{} \gamma_\mu (1-\gamma_5) b \,
		  \ov c{}  \gamma^\mu \gamma_5 c,
\nn
Q_{8V} & \equiv & \ov s{} \gamma_\mu (1-\gamma_5) T^a b\,
		  \ov c{}  \gamma^\mu T^a c,
		\nn
Q_{8A} & \equiv & \ov s{} \gamma_\mu (1-\gamma_5) T^a b\,
		  \ov c{}  \gamma^\mu \gamma_5 T^a c.
\label{eq:ops}
\end{eqnarray}
Then $T_f$ in \eq{eq:tp} becomes $T_f= \frac{G_F}{\sqrt 2}
\braOket{J/\psi K_S}{C_0(Q_{0V}-Q_{0A}) +C_8 (Q_{8V}-Q_{8A})}{B_d}$ with
$C_0 = C_2/N_c +C_1$ and $C_8=2C_2$, where $N_c=3$ is the number of
colors.  Using next-to-leading order (NLO) Wilson coefficients in the
\uli{naive dimensional regularization} (NDR) scheme \cite{bw,bbl} at the
scale $\mu=m_{\psi}$ one finds $C_0=0.13 $ and $C_8=2.2$. The smallness
of $C_0$ is a well-known numerical accident entailing that the weak
decay produces the $(c,\ov c)$ pair almost in a color octet state.  We
normalize the matrix elements (for $j=0,8$) as
\begin{equation}
 \langle Q_{jV} \rangle =
V_0
\, v_j, \qquad
 \langle Q_{jA} \rangle =
V_0 \, a_j 
\label{eq:v0aj}
\end{equation}
to the factorized matrix element $ V_0\equiv \langle Q_{0V} \rangle {}_{\rm fact}
=2 f_{J/\psi} m_{B_d} p_{cm} F_1^{B\to K}(m_{\psi}^2)= (4.26 \pm 0.16)
\gev{}^{3} $.
The uncertainty stems from the form factor $F_1^{B\to
  K}(m_{\psi}^2)=0.586 \pm 0.021$ \cite{ff} and the $J/\psi$ decay
constant $f_{J/\psi}=(0.405\pm 0.005) \gev$.  $m_{B_d}=5.28\gev$ and
$p_{cm}=1.68 \gev$
are the $B_d$ mass and the three-momentum of the $J/\psi$ or $K_S$ in
the $B_d$ rest frame. $v_{0,8},a_{0,8}$ depend on $\mu$ in such a way
that the $\mu$-dependence of $C_0,C_8$ cancels from physical
quantities. When we quote numerical values we refer to the choice
$\mu=m_{\psi}$.  The large-$N_c$ counting of our (complex) hadronic
parameters is $v_0=1+{\cal O}(1/N_c^2)$, $v_8,a_8={\cal O}(1/N_c)$, and
$a_0={\cal O}(1/N_c^2)$. Normalizing the branching ratio to the
experimental value we find
\begin{eqnarray}
  \frac{B(B_d\to J/\psi K_S)}{B(B_d\to J/\psi K_S)_{\rm exp}}
  &=& \nn 
&& \hspace{-15mm} 
  \lt[1\pm 0.08 \rt] \lt| 0.47v_0 +7.8 (v_8-a_8) \rt|^2.
\label{eq:bf}
\end{eqnarray}
\begin{figure*}[t]
	\begin{subfigure}{0.23\textwidth}
		\includegraphics[width=\textwidth,clip=true]
 {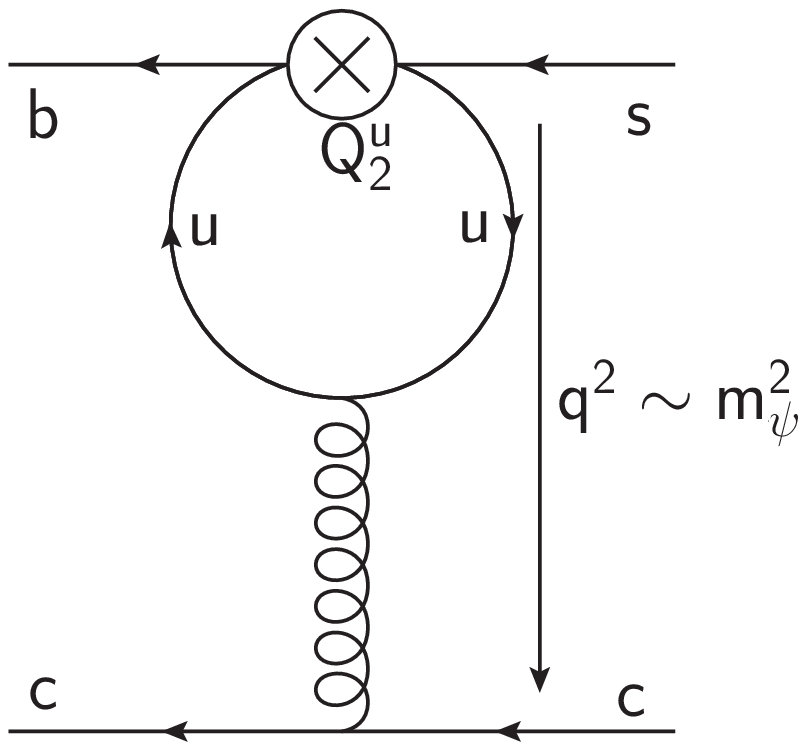}
		\caption{LO Penguin \label{fig:1a}}
		
	\end{subfigure}
	\begin{subfigure}{0.2\textwidth}
		\includegraphics[width=\textwidth,clip=true]%
		  {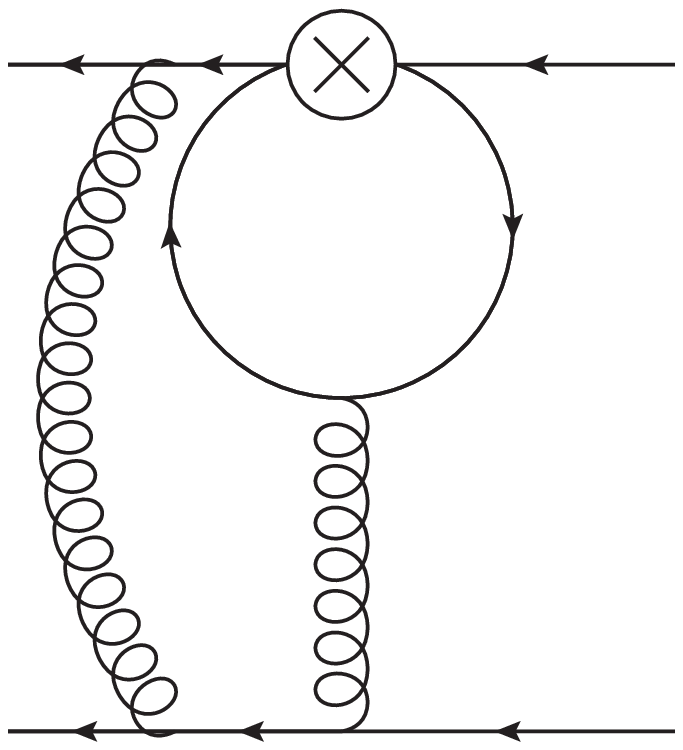}
		  \caption{\label{fig:1b}} 
		   
	\end{subfigure}
	\begin{subfigure}{0.15\textwidth}
		\includegraphics[width=\textwidth,clip=true]%
		  {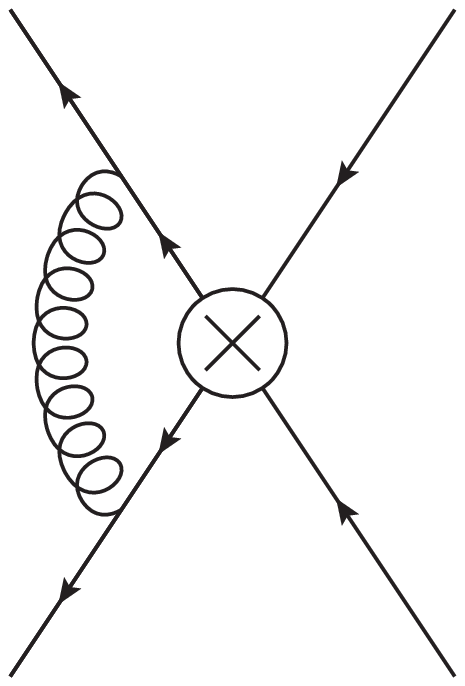}
		  \caption{\label{fig:1c}	} 
		  	  
	\end{subfigure}
	\begin{subfigure}{0.2\textwidth}
		\includegraphics[width=\textwidth,clip=true]%
		   {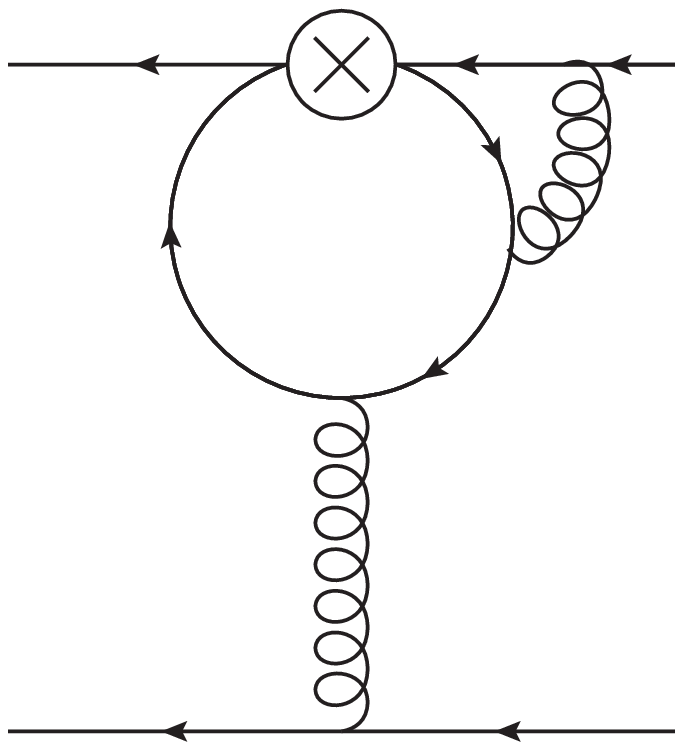}
		  \caption{}
		  \label{fig:1d}		  
	\end{subfigure}
	\begin{subfigure}{0.19\textwidth}
		\includegraphics[width=\textwidth,clip=true]%
		      {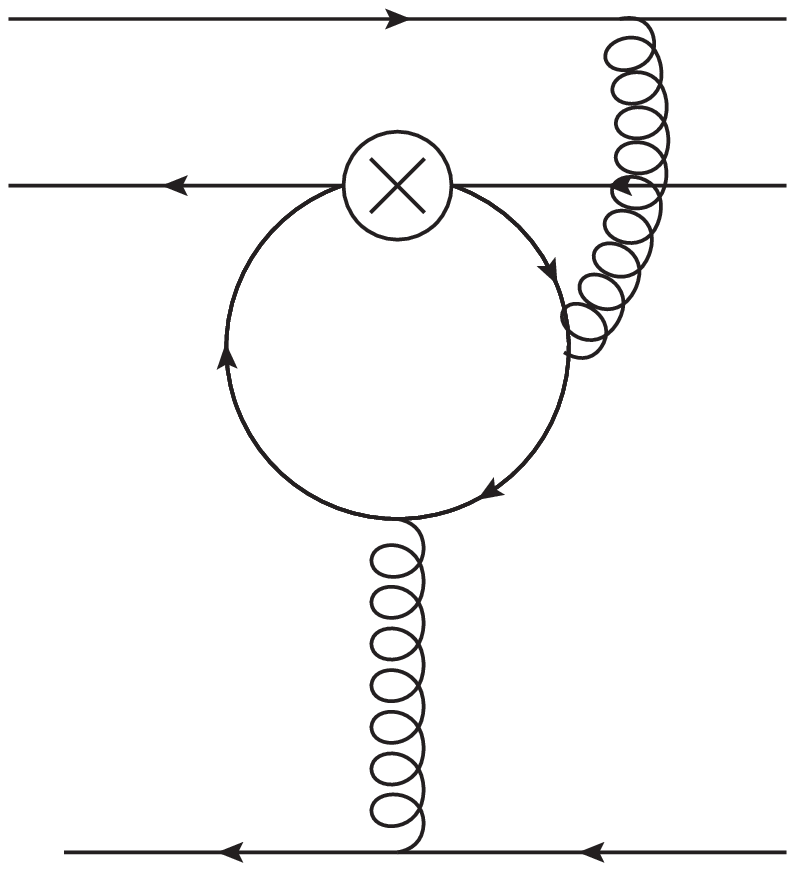}
		         \caption{Spectator scattering}
		         \label{fig:1e}
	\end{subfigure}
  \caption{The LO diagram is shown in (a). The soft IR divergence of the
    diagram (b) factorizes with the corresponding diagram 
    of the effective-theory side shown in (c).  
    The diagram (d) is an example of a diagram with a collinear IR 
    divergence.  In (e) a spectator diagram is given.}\label{fig:pengs}
~\\[-3mm]
\hrule
\end{figure*}
Varying the phase of $v_8-a_8$ between $-\pi$ and $\pi$ one finds the
correct branching ratio for $0.07\leq |v_8-a_8|\leq 0.19$ if $v_0$ is
set to 1.  Thus, there is no mystery with the branching ratio and the
hadronic parameters obey the hierarchy expected from $1/N_c$ counting.
The terms involving $a_0$ are negligible in view of other uncertainties
and are omitted throughout this paper.

$P_f$ in \eq{eq:tp} receives contributions from $Q_{1,2}^u$ and the
penguin operators $Q_j$, $j\geq 3$. The matrix elements of the latter
can be trivially expressed in terms of the operators in
\eq{eq:ops}. Therefore, this contribution to $P_f/T_f$ only depends on
$v_8/v_0$ and $a_8/v_0$. Below we will see that the magnitudes
of these ratios are under control thanks to the $1/N_c$ hierarchy of
$v_0,v_8,a_8$ and the information from $B(B_d\to J/\psi K_S)_{\rm exp}$.
By varying the parameters in the allowed ranges we can then find the
maximal contribution of the penguin operators to
$|\Delta\phi|$. 

In order to apply the same strategy to $Q_{1,2}^u$ we
must first express the up-quark penguin depicted in 
\fig{fig:1a} in terms of matrix elements of the local operators in
\eq{eq:ops}. In Ref.~\cite{Bander:1979px} it is argued that a penguin
loop flown through by a hard momentum $q$ (in our case $q^2\sim
m_\psi^2=(3.1\gev)^2$) can be calculated in perturbation theory
(``\uli{Bander-Soni-Silverman (BSS)}
mechanism'').  In Ref.~\cite{Boos:2004xp} this idea is used to find an
estimate of $\langle Q_2^u\rangle$ which leads to an upper bound on
$|\Delta\phi|$ which is smaller than the values found by SU(3)$_{\rm F}$
arguments \cite{su3}. In this paper, we turn the BSS idea into a rigorous
field-theoretic method by proving an operator product expansion (OPE)
\begin{eqnarray}
\!\!  \braOket{J/\psi K_S}{Q_j^u}{B_d} &=&  \sum_k
     \widetilde{C}_{j,k}
  \braOket{J/\psi K_S}{Q_{k}}{B_{d}}
    + \ldots 
\label{eq:ope}
\end{eqnarray}
with $k$ running over $k=0V,0A,8V,8A$. 
The dots representing terms suppressed by higher powers of
$\lqcd/\sqrt{q^2}$.  The Wilson coefficients
$\widetilde{C}_{j,k}=\widetilde{C}_{j,k}^{(0)}+
(\alpha_s(\mu)/(4\pi))\widetilde{C}_{j,k}^{(1)}+\ldots $ are calculated
in perturbation theory to the desired order in $\alpha_s(\mu)$, with the
renormalization scale $\mu={\cal O}(m_{\psi},m_b)$. A similar OPE has
been derived to calculate charm-loop effects in the rare semileptonic
decays $B\to K^{(*)} \ell^+\ell^-$ \cite{rare}. Since leptons carry no
color charges, this application involves no four-quark operators like
those in \eqsand{eq:ops}{eq:ope}. From \fig{fig:1a} one finds $
\widetilde{C}_{j,k}^{(0)}=0$ except for
$\widetilde{C}_{8G,8V}^{(0)}=-\frac{m_b^2}{q^2}\frac{\alpha_s}{\pi}$ and
$\widetilde{C}_{2,8V}^{(0)} =P(q^2) $ with the penguin function
\begin{eqnarray}
P(q^2)& = &\frac23 \frac{\alpha_s}{4 \pi}\lt[ \ln\lt(\frac{q^2}{\mu^2}\rt) -i 
\pi -\frac23 \rt].
\end{eqnarray}
{Inherent to applications of the OPE as in Ref.~\cite{rare} or in this
  paper is the assumption that rescattering effects for values of $q^2$
  far above the partonic pair-production threshold are correctly
  described in perturbation theory. \eq{eq:ope} captures all hadronic
  effects in the $(u,\ov u)\rightarrow (c,\ov c)$ transition only if
  there is no intrinsic $(u,\ov u)$ component in the $J/\psi$ wave
  function (e.g.\ no $J/\psi$--$\rho^0$ mixing).  A powerful check of our
  framework will be the confrontation of our predictions for $b\to c\ov
  c d$ transitions with data.}

\section{Proof of Factorization}
In order to establish \eq{eq:ope} we must prove that the coefficients $
\widetilde{C}_{j,k}$ are infrared (IR) safe. To this end we analyze {i)}
the soft IR divergences of the two-loop diagrams contributing to
$\langle Q_j^u\rangle$,
{ii)} the collinear IR divergences of these diagrams, {iii)}
spectator scattering diagrams, and {iv)} higher-order diagrams in
which the large momentum bypasses the penguin loop (``long distance
penguins'').

{i)} An example of a diagram with a soft divergence is shown in
\fig{fig:1b}. This soft divergence is reproduced by the corresponding
diagram of the effective-theory side (i.e.\ RHS) of \eq{eq:ope},
depicted in \fig{fig:1c}, so that this divergence
factorizes with $\widetilde{C}_{j,k}^{(0)}$ and does not affect
$\widetilde{C}_{j,k}^{(1)}$. All soft divergences are from diagrams in
which the additional gluon connects two external lines and cancel from
$\widetilde{C}_{j,k}^{(1)}$ in the same way. 

{ii)} Collinear divergences occur in diagrams in which a gluon is
attached to the line with the strange quark, which we treat as
massless. An example is shown in \fig{fig:1d}. If $l$ denotes the loop
momentum flowing through the gluon propagator and $p_s$ is the momentum
of the external strange quark, the collinear divergence corresponds to
the region with $l^2=0$ and $l \propto p_s$. We can then reduce the
problem to the study of one-loop diagrams with an external on-shell
gluon: If we sum over all possibilities to attach this gluon to one of
the lines of the LO diagram in \fig{fig:1a}, the collinear Ward identity
of QCD ensures that this sum vanishes when the open Lorentz index of the
gluon line is contracted with $l^\mu$. This feature ensures that the
collinear divergences of the sum of the two-loop diagrams vanish. (For a
discussion in the context of QCD factorization see
Refs.~\cite{qcdf,Bauer:2001cu,Buchalla:2002pd}.)  It equally holds for
the effective-theory side of the OPE. 
The cancellation of collinear divergences is conceptually identical to
the situation in typical processes in collider physics; it is further
known to be much simpler (with fewer diagrams to be discussed) if a
physical gauge (with only two propagating gluon degrees of freedom) is
adopted.

{iii)} Next we discuss the spectator scattering contributions: diagrams
in which the gluon connects the $b$ or $s$ line with the spectator quark
line trivially factorize with the corresponding diagrams on the
effective side.  If the gluon {connects} the {spectator} with the gluon
{line or a} charm or up line, we have to take into account that the
squared momentum in the penguin loop is $(q+l)^2$ instead of $q^2$. If
the gluon is soft, $l^\mu\sim\lqcd$, the expansion of the loop function
$P$ around $q^2$ reproduces a term which correctly factorizes with
$\widetilde{C}_{j,k}^{(0)}$ up to term suppressed by
$\lqcd/\sqrt{q^2}$. If the gluon is hard-collinear, with virtuality
$l^2\sim p_{\rm cm} \lqcd$, {where $p_{\rm cm}\sim 1.5\gev$ is the
  three-momentum of the $K_S$ or $J/\psi$ in the $B_d$ rest frame,} the
situation is more subtle: the LO diagram is suppressed by $\lqcd/p_{\rm
  cm}$, because the momentum of the spectator quark changes from zero to
${\cal O}(p_{\rm cm})$ in the decay, which is penalized by the
light-cone distribution amplitude (LCDA) of the kaon \cite{qcdf}. The
asymptotic form of the kaon LCDA, $\Phi(x)=6x(1-x)$, where $x$ and $1-x$
are the fractions of the kaon momentum carried by the $\ov s$ and $d$
quarks, {favors} momentum configurations in which the kaon momentum is
roughly equally shared between the two valence quarks. While the
propagator of the scattered hard-collinear gluon is suppressed as $\sim
1/(\lqcd p_{\rm cm})$, the suppression of the LO diagram is lifted,
because the spectator momentum is in the region $x\sim 1/2$ {favored} by
the kaon LCDA. To identify further suppression factors we first discuss
the case that the gluon connect a charm line with the spectator:
counting $q^2\sim m_{\psi}^2$ and the energies of $\ov s$ and
spectator-$d$ quarks as $p_{\rm cm}/2$, the penguin loop gives
$P((q+l)^2)\simeq P(q^2) + \frac{p_{\rm cm}} {m_{\psi}}P^\prime(q^2)$.
The non-factorizing piece involving the derivative $P^\prime(q^2)$
{comes with a factor of $p_{\rm cm}/m_{\psi}$.}  The virtuality of
the (anti-)charm propagator is around $p_{\rm cm}$ {entailing a
  suppression factor of $\lqcd/p_{\rm cm}$}.  Thus, \uli{altogether}
spectator scattering from the charm lines obeys \eq{eq:ope} up to terms
of order $\lqcd/m_{\psi}$.  Next we discuss the spectator scattering
from the up line, with a sample diagram depicted in \fig{fig:1e}. We
find that these diagrams are power-suppressed by $\lqcd/m_{\psi}$. In
this respect these spectator diagrams differ from the similar photon
penguins calculated in Ref.~\cite{Beneke:2006mk}, which involve $P(q^2)$
for $q^2\sim 0$ rather than $q^2\sim m_{\psi}^2$.

{vi)} So far we have assumed
that the underlying hard process is the penguin loop with the hard scale
$\sqrt{q^2}$. But it may also be possible that the hard momentum
transfer to the $J/\psi$ occurs through a hard gluon radiated from the
$b$ or $s$ line, while the penguin loop is {a ``long-distance
  penguin''} governed by soft QCD. Such a situation is exemplified by
the diagram in \fig{fig:1b} with the left gluon having virtuality $\sim
m_{\psi}^2$. These diagrams, in which the whole weak decay process
occurs with small momentum transfers, have a suppression factor
$(\lqcd/\sqrt{q^2})^3$ stemming from the hard gluon propagator and an
off-shell $b$ quark propagator (or $s$ quark propagator).

In our power counting {in i)--iv)} we have treated $p_{\rm cm}$ as an
intermediate scale between $\lqcd$ and $m_\psi$ and have found no
non-factorizable non-perturbative effects of order $p_{\rm
  cm}/m_\psi$. While $p_{\rm cm}$ enters two-loop diagrams through
$p_b\cdot p_s \sim m_bp_{\rm cm} $, such terms do not come with IR
divergences and end up in the NLO corrections to the coefficients
$\widetilde{C}_{j,k}$. We find that the counting rule for $p_{\rm cm}$
is irrelevant, one can reproduce our results above as well for the
limiting cases $p_{\rm cm}\sim \lqcd$ and $p_{\rm cm}\sim \sqrt{q^2}$.
{In particular, higher orders of the OPE do not involve operators
  with derivatives acting on the $\ov s$ field. The same feature was found for
  $B\to K^{(*)}\ell^+\ell^-$ in the last paper of Ref.~\cite{rare}.}

The choice $q^2=m_\psi^2$ for $P(q^2)$ may be altered by adding a 
contribution of order $\lqcd$ to $\sqrt{q^2}$. This shuffles a piece 
proportional to $(\lqcd/m_\psi) P^\prime(m_\psi^2) $ into the
coefficient of the sub-leading operator $\ov s{} \gamma_\mu (1-\gamma_5) T^a b\,
\lt[ \Box-m_\psi^2 \rt] \ov c{}  \gamma^\mu T^a c $, which removes the
ambiguity associated with the choice of $q^2$. At NLO in  $\alpha_s$ 
one generates non-zero coefficients $\widetilde{C}_{j,k}^{(1)}$ also for 
$j=1$ or $k=0A,8A$. 

In conclusion the OPE with the minimal set of operators in \eq{eq:v0aj}
works, the coefficients in $\widetilde{C}_{j,k}$ are IR-safe.

\section{Phenomenology}
The penguin amplitude {depends on the Wilson coefficients as}
\begin{equation}
  P_f= V_0 \lt( 2 C_4 +2 C_6 
  +2 C_2 
  \widetilde{C}_{2,8V}^{(0)}+C_{8G} \widetilde{C}_{8G,8V}^{(0)}\rt)v_8
    + \dots 
  \label{eq:24p}
\end{equation}
where the dots represent the terms with $v_0$ and $a_8$ which have
  much smaller coefficients. The dependence of $\widetilde{C}_{2,8V}^{(0)}$
  (calculated from \fig{fig:1a}) on the renormalization scheme cancels
  with the scheme dependence of $C_4+C_6$ in \eq{eq:24p}. In the NDR
  scheme adopted by us these penguin coefficients give a larger 
  contribution to $P_f$ than the $u$-penguin loop contained in
  $\widetilde{C}_{2,8V}^{(0)}$. This is not surprising, because
    the  $u$-penguin loop enters at NLO, while $C_4+C_6$ already
    contributes at LO. The omission of this dominant LO piece explains
    the smallness of the result in Ref.~\cite{Boos:2004xp}.

  For the prediction of $P_f/T_f$ we implement the constraint from
  $B(B\to f)$ exemplified for $f={J/\psi K_S}$ in \eq{eq:bf} in the
  following way: adapting a phase convention in which $A_f$ in
  \eq{eq:atp} is {real and} positive, we can determine $a_8$ in terms of
  {$V_0$,} $v_0$, $v_8$, and the measured $B(B\to f)$
  \uli{\cite{BranchingRatios,formfactors}}.  Then we use this to eliminate $a_8$ from
  $P_f/T_f$.  For example, we find
\begin{equation}
\frac{P_{J/\psi K_S}}{T_{J/\psi K_S}}= 0.01- 0.02 v_0 - (0.71 +0.33i)v_8 
\label{eq:ptv}
\end{equation}
{for} the central \uli{value} of $V_0$ {(quoted after \eq{eq:v0aj})}
and $B(B_d \to J/\psi K_S)$. {We vary $v_8$ and $v_0$ in their
  allowed ranges $|v_8|<1/3$ and $|v_0|=1\pm 0.15$ 
  with the constraint that $|a_8|\leq 
  1/3$ 
  must be obeyed. The
  allowed ranges for $\Delta \phi$, $C_f$, and $\Delta S_f \equiv S_f +
  \eta_f \sin \phi_q$ are almost symmetric around zero. We list the
  upper bounds on their magnitudes 
  {for {several} decay modes} in Tabs.~\ref{tab:resultsBd} and
  \ref{tab:resultsBs}. The results include the uncertainties from $V_0$,
  the branching ratios, CKM parameters \cite{Charles:2004jd}, 
  and higher-order terms in our 
  OPE.  For the
  $b\to \ov{c} c d$ decay modes with Cabibbo-unsuppressed $P_f/T_f$ the
  expansion in \eq{eq:Dphi} has been replaced by the exact formula (see
  e.g.\ Ref.~\cite{Fleischer:1999nz,su3}). Our bounds are conservative, as
  the considered ranges for $v_8$ and $a_8$ are wide (permitting
  even sizable cancellations in \eq{eq:bf}).  From
  \eqsand{eq:bf}{eq:ptv} one verifies that any additional information on
  magnitude or phase of one of these parameters
  will substantially reduce the ranges quoted in
  Tabs.~\ref{tab:resultsBd} and \ref{tab:resultsBs}. Our results for
  $B_d\to J/\psi \pi^0$ favor the Belle measurement $C_{J/\psi
    \pi^0}=-0.08\pm0.17$, $S_{J/\psi \pi^0}=-0.65\pm0.22$ \cite{Lee:2007wd} 
  over the BaBar
  result $C_{J/\psi \pi^0}=-0.20\pm0.19$, $S_{J/\psi \pi^0}=-1.23\pm0.21$}
\cite{Aubert:2008bs}. (In the absence of penguin pollution $C_{J/\psi
  \pi^0}=0$ and $S_{J/\psi \pi^0}=-\sin(2\beta)=-0.69\pm 0.02$.)
In the the case of a more precise and non-vanishing measurement of 
$C_{J/\psi \pi^0}$, for example, $C_{J/\psi \pi^0}=-0.10\pm0.01$, which corresponds
to the current world average with a ten times smaller error, we can also
put stronger restrictions on the shift of the mixing-induced 
\CP violation $|\Delta S_{J/\psi \pi^0}|\leq 0.13 $.
A measurement of $C_{J/\psi \pi^0}$ that is consistent with zero,
however, does not improve the bound. This {feature occurs in} all
decay modes {with} Cabibbo-unsuppressed {$P_f/T_f$. The
  measurements of $S_f$ and $C_f$ for the $B_d\to J/\psi \rho^0$
  polarization amplitudes \cite{Aaij:2014vda} comply with the ranges in 
  Tab.~\ref{tab:resultsBd}.} 
 
\begin{table*}
  \caption{The maximal phase shift of $\phi_d$ due to penguin 
    pollution and limits for the $\CP$ violation observables $S_f$ and $C_f$ in 
    various $B_d\to f$ decays. 
    {Decays into two vector mesons involve different polarization
      amplitudes, 
  indicated by $0$, $\parallel$, and $\perp$ \cite{Dighe:1995pd}.
  In $S_f$ for $f=J/\psi K^*$ $K^*\to K_s\pi^0$ is understood. } 
}

\begin{ruledtabular}
\begin{tabular}{lccccccccc}
  Final State \rule{0pt}{2.3ex} & $J/\psi K_S$ & $\psi(2S) K_S$ & 
  $J/\psi\pi^0$  & $(J/\psi\rho)^0$&  $(J/\psi\rho)^\parallel$ & 
  $(J/\psi\rho)^\perp$& $(J/\psi K^*)^0$&$ (J/\psi K^*)^\parallel$& $(J/\psi 
  K^*)^\perp $\\\hline
  {$\max(|\Delta \phi_d|)$  $[{}^\circ]$}&{$0.68$}&{$0.74$ }&{ n. a.  }&{ n. a.  }&{ n. a.  }&{ n. a. 
  }&{$0.85$}&{$1.13$}&{$0.93$} 
    \\ 
    $\max(|\Delta S_f|)$ $[10^{-2}]$&{$0.86$}&{$0.94$}&{$18.$}&{$22.$}&{$27.$}&{$22.$}&{$1.09$}&{$1.45$ }&{$1.19$} \\ 
  $\max(|C_f|)$ $[10^{-2}]$&{$1.33$}&{$1.33$}&{$29.$}&{$35.$}&{$41.$}&{$36.$}&{$1.65$}&{$2.19$}&{$1.80$
}
\end{tabular}
\end{ruledtabular}
\label{tab:resultsBd}
\end{table*}

\begin{table*}
\caption{{Same as Tab.~\ref{tab:resultsBd} for  $B_s\to f$ decays.}}
\begin{ruledtabular}
\begin{tabular}{lccccccc}
  Final State & $J/\psi K_S$&   $(J/\psi\phi)^0$& 
  $(J/\psi\phi)^\parallel$& $\left(J/\psi\phi\right)^\perp$& $(J/\psi K^*)^0$&$ 
  (J/\psi K^*)^\parallel$& $(J/\psi K^*)^\perp$  \\\hline
  {
    $\max(|\Delta \phi_s|)$  $[{}^\circ]$}&{n.a.}&{$0.97$}&{$1.22$}&{$0.99$}&{n.a.}&{n.a.}&{n.a.}\\ 
    $\max(|\Delta S_f|)$ $[10^{-2}]$&{$26.$}&{$1.70$}&{$2.13$}&{$1.73$}&{$40.$}&{$58.$}&{$35.$} \\ 
      $\max(|C_f|)$ $[10^{-2}]$&{$27.$}&{$1.89$}&{$2.35$}
      &{$1.92$}&{$43.$}&{$64.$}&{$37.$}
 \end{tabular}
\end{ruledtabular}
\label{tab:resultsBs}
\end{table*}

\section{Conclusions}
We have established a factorization formula (to leading power in
$\lqcd/m_{\psi}$) for the penguin contribution to the
\CP$\!\!$-violating coefficients $S_f$ and $C_f$ in $A_{\rm CP}^{B_q\to
  f}(t)$ for final states $f$ containing charmonium and the related
shift $\Delta \phi_q$ {of the corresponding \CP phase}. As a crucial
result the penguin contributions involve the same hadronic matrix
elements as the tree amplitude. This allows us to constrain $P_f/T_f$,
which determines $S_f$ and $C_f$, and to find e.g.\ $|\Delta \phi_d
|\leq {0.68}^\circ$ for $B_d\to J/\psi K_S$ and $|\Delta
\phi_s^\perp|\leq {0.99}^\circ $ for $B_d\to J/\psi\phi$, representing
bounds that were thought to be uncalculable from first principles. Novel
territory are our predictions for $S_f$ and $C_f$ in $b\to c\ov c d$
decays, in which $P_f/T_f$ is Cabibbo-unsuppressed. {Future experimental
  probes of these predictions will constitute a powerful test of our
  theoretical framework, {whose key ingredient is an operator
    product expansion for the up-quark penguin loop.}}  There
are no similar consistency checks for the standard predictions of
$P_f/T_f$ based on SU(3)$_F$ symmetry, which, moreover, cannot be used
for $B_s\to J/\psi \phi$. We further remark that our results do not
depend on any properties of the charmonium {LCDA}.

\section{Acknowledgments}
We thank Gerhard Buchalla, Marco Ciuchini, Enrico Franco, Yuval
Grossman, Sebastian J\"ager, Kirill Melnikov, and Luca Silvestrini for
fruitful discussions .  This work is supported by BMBF under grant
no.~05H12VKF.  P.F.~acknowledges the support by the DFG-funded Doctoral
School \emph{KSETA}.

\end{document}